\documentclass[prc,aps,twocolumn,showpacs,nofootinbib]{revtex4}
\pdfoutput=1 
\usepackage{amsmath}
\usepackage{amsfonts}
\usepackage{graphicx}
\usepackage{dcolumn}
\usepackage{hyperref}

\begin{document}


\newenvironment{ibox}[1]%
{\vskip 1.0em
\framebox[\columnwidth][r]{%
\begin{minipage}[c]{\columnwidth}%
\vspace{-1.0em}%
#1%
\end{minipage}}}
{\vskip 1.0em}

\newcommand{\iboxed}[1]{%
\vskip 1.0em
\framebox[\columnwidth][r]{%
\begin{minipage}[c]{\columnwidth}%
\vspace{-1.0em}
#1%
\end{minipage}}
\vskip 1.0em}

\newcommand{\fitbox}[2]{%
\vskip 1.0em
\begin{flushright}
\framebox[{#1}][r]{%
\begin{minipage}[c]{\columnwidth}%
\vspace{-1.0em}
#2%
\end{minipage}}
\end{flushright}
\vskip 1.0em}

\newcommand{\iboxeds}[1]{%
\vskip 1.0em
\begin{equation}
\fbox{%
\begin{minipage}[c]{1mm}%
\vspace{-1.0em}
#1%
\end{minipage}}
\end{equation}
\vskip 1.0em}

\def\Xint#1{\mathchoice
   {\XXint\displaystyle\textstyle{#1}}%
   {\XXint\textstyle\scriptstyle{#1}}%
   {\XXint\scriptstyle\scriptscriptstyle{#1}}%
   {\XXint\scriptscriptstyle\scriptscriptstyle{#1}}%
   \!\int}
\def\XXint#1#2#3{{\setbox0=\hbox{$#1{#2#3}{\int}$}
     \vcenter{\hbox{$#2#3$}}\kern-.5\wd0}}
\def\ddashint{\Xint=}
\def\dashint{\Xint-}

\newcommand{\alps}{\ensuremath{\alpha_s}}
\newcommand{\qbar}{\bar{q}}
\newcommand{\ubar}{\bar{u}}
\newcommand{\dbar}{\bar{d}}
\newcommand{\sbar}{\bar{s}}
\newcommand{\beq}{\begin{equation}}
\newcommand{\eeq}{\end{equation}}
\newcommand{\beqa}{\begin{eqnarray}}
\newcommand{\eeqa}{\end{eqnarray}}
\newcommand{\gs}{g_{\pi NN}}
\newcommand{\gw}{f_\pi}
\newcommand{\mq}{m_Q}
\newcommand{\mn}{m_N}
\newcommand{\mpi}{m_\pi}
\newcommand{\mrho}{m_\rho}
\newcommand{\momg}{m_\omega}
\newcommand{\bb}{\langle}
\newcommand{\kb}{\rangle}
\newcommand{\xvec}{\mathbf{x}}
\newcommand{\st}{\ensuremath{\sqrt{\sigma}}}
\newcommand{\Bvec}{\mathbf{B}}
\newcommand{\rvec}{\mathbf{r}}
\newcommand{\kvec}{\mathbf{k}}
\newcommand{\bvec}[1]{\ensuremath{\mathbf{#1}}}
\newcommand{\bra}[1]{\ensuremath{\bb#1|}}
\newcommand{\ket}[1]{\ensuremath{|#1\kb}}
\newcommand{\gft}{\ensuremath{\gamma_{FT}}}
\newcommand{\gfv}{\ensuremath{\gamma_5}}
\newcommand{\bfalp}{\ensuremath{\bm{\alpha}}}
\newcommand{\bfbeta}{\ensuremath{\bm{\beta}}}
\newcommand{\bfeps}{\ensuremath{\bm{\epsilon}}}
\newcommand{\lag}{{\lambda_\gamma}}
\newcommand{\lao}{{\lambda_\omega}}
\newcommand{\lN}{\lambda_N}
\newcommand{\lM}{\lambda_M}
\newcommand{\lB}{\lambda_B}
\newcommand{\epslag}{\ensuremath{\bm{\epsilon}_{\lag}}}
\newcommand{\bfept}{\ensuremath{\tilde{\bm{\epsilon}}}}
\newcommand{\bfgam}{\ensuremath{\bm{\gamma}}}
\newcommand{\bfnab}{\ensuremath{\bm{\nabla}}}
\newcommand{\bflambda}{\ensuremath{\bm{\lambda}}}
\newcommand{\bfmu}{\ensuremath{\bm{\mu}}}
\newcommand{\bfphi}{\ensuremath{\bm{\phi}}}
\newcommand{\bfvphi}{\ensuremath{\bm{\varphi}}}
\newcommand{\bfpi}{\ensuremath{\bm{\pi}}}
\newcommand{\bfsig}{\ensuremath{\bm{\sigma}}}
\newcommand{\bftau}{\ensuremath{\bm{\tau}}}
\newcommand{\bfrho}{\ensuremath{\bm{\rho}}}
\newcommand{\bfth}{\ensuremath{\bm{\theta}}}
\newcommand{\bfchi}{\ensuremath{\bm{\chi}}}
\newcommand{\bfxi}{\ensuremath{\bm{\xi}}}
\newcommand{\bfR}{\ensuremath{\bvec{R}}}
\newcommand{\bfP}{\ensuremath{\bvec{P}}}
\newcommand{\Rcm}{\ensuremath{\bvec{R}_{CM}}}
\newcommand{\spup}{\uparrow}
\newcommand{\spd}{\downarrow}
\newcommand{\up}{\uparrow}
\newcommand{\dn}{\downarrow}
\newcommand{\hbarom}{\frac{\hbar^2}{m_Q}}
\newcommand{\half}{\ensuremath{\frac{1}{2}}}
\newcommand{\thalf}{\ensuremath{\frac{3}{2}}}
\newcommand{\fhalf}{\ensuremath{\frac{5}{2}}}
\newcommand{\shalf}{\ensuremath{{\tfrac{1}{2}}}}
\newcommand{\sqtr}{\ensuremath{{\tfrac{1}{4}}}}
\newcommand{\sphalf}{\ensuremath{\genfrac{}{}{0pt}{1}{+}{}\!\tfrac{1}{2}}}
\newcommand{\smhalf}{\ensuremath{\genfrac{}{}{0pt}{1}{-}{}\!\tfrac{1}{2}}}
\newcommand{\sthalf}{\ensuremath{{\tfrac{3}{2}}}}
\newcommand{\spthalf}{\ensuremath{{\tfrac{+3}{2}}}}
\newcommand{\smthalf}{\ensuremath{{\tfrac{-3}{2}}}}
\newcommand{\sfhalf}{{\tfrac{5}{2}}}
\newcommand{\third}{{\frac{1}{3}}}
\newcommand{\tthird}{{\frac{2}{3}}}
\newcommand{\sthird}{{\tfrac{1}{3}}}
\newcommand{\stthird}{{\tfrac{2}{3}}}
\newcommand{\vnn}{\ensuremath{\hat{v}_{NN}}}
\newcommand{\vij}{\ensuremath{\hat{v}_{ij}}}
\newcommand{\vik}{\ensuremath{\hat{v}_{ik}}}
\newcommand{\argonne}{\ensuremath{v_{18}}}
\newcommand{\lqcd}{\ensuremath{\mathcal{L}_{QCD}}}
\newcommand{\lqed}{\ensuremath{\mathscr{L}_{QED}}}
\newcommand{\lgf}{\ensuremath{\mathcal{L}_g}}
\newcommand{\lqm}{\ensuremath{\mathcal{L}_q}}
\newcommand{\lqg}{\ensuremath{\mathcal{L}_{qg}}}
\newcommand{\nn}{\ensuremath{N\!N}}
\newcommand{\nnn}{\ensuremath{N\!N\!N}}
\newcommand{\qq}{\ensuremath{qq}}
\newcommand{\qqq}{\ensuremath{qqq}}
\newcommand{\qqb}{\ensuremath{q\bar{q}}}
\newcommand{\hpnd}{\ensuremath{H_{\pi N\Delta}}}
\newcommand{\hpqq}{\ensuremath{H_{\pi qq}}}
\newcommand{\hpqqa}{\ensuremath{H^{(a)}_{\pi qq}}}
\newcommand{\hpqqe}{\ensuremath{H^{(e)}_{\pi qq}}}
\newcommand{\hint}{\ensuremath{H_{\rm int}}}
\newcommand{\fpnn}{\ensuremath{f_{\pi\! N\!N}}}
\newcommand{\fenn}{\ensuremath{f_{\eta\! N\!N}}}
\newcommand{\gsnn}{\ensuremath{g_{\sigma\! N\!N}}}
\newcommand{\gpnn}{\ensuremath{g_{\pi\! N\!N}}}
\newcommand{\fpnd}{\ensuremath{f_{\pi\! N\!\Delta}}}
\newcommand{\grpg}{\ensuremath{g_{\rho\pi\gamma}}}
\newcommand{\gopg}{\ensuremath{g_{\omega\pi\gamma}}}
\newcommand{\fmqq}{\ensuremath{f_{M\! qq}}}
\newcommand{\gmqq}{\ensuremath{g_{M\! qq}}}
\newcommand{\fpqq}{\ensuremath{f_{\pi qq}}}
\newcommand{\gpqq}{\ensuremath{g_{\pi qq}}}
\newcommand{\feqq}{\ensuremath{f_{\eta qq}}}
\newcommand{\gonn}{\ensuremath{g_{\omega N\!N}}}
\newcommand{\gonna}{\ensuremath{g^t_{\omega N\!N}}}
\newcommand{\grnn}{\ensuremath{g_{\rho N\!N}}}
\newcommand{\gopr}{\ensuremath{g_{\omega\pi\rho}}}
\newcommand{\grnp}{\ensuremath{g_{\rho N\!\pi}}}
\newcommand{\grpp}{\ensuremath{g_{\rho\pi\pi}}}
\newcommand{\Lpnn}{\ensuremath{\Lambda_{\pi\! N\! N}}}
\newcommand{\Lonn}{\ensuremath{\Lambda_{\omega N\! N}}}
\newcommand{\Lonna}{\ensuremath{\Lambda^t_{\omega N\! N}}}
\newcommand{\Lrnn}{\ensuremath{\Lambda_{\rho N\! N}}}
\newcommand{\Lopr}{\ensuremath{\Lambda_{\omega\pi\rho}}}
\newcommand{\Lrpp}{\ensuremath{\Lambda_{\rho\pi\pi}}}
\newcommand{\getaqq}{\ensuremath{g_{\eta qq}}}
\newcommand{\fsqq}{\ensuremath{f_{\sigma qq}}}
\newcommand{\gsqq}{\ensuremath{g_{\sigma qq}}}
\newcommand{\piqq}{\ensuremath{{\pi\! qq}}}
\newcommand{\ylm}{\ensuremath{Y_\ell^m}}
\newcommand{\ylmc}{\ensuremath{Y_\ell^{m*}}}
\newcommand{\ebh}[1]{\hat{\bvec{e}}_{#1}}
\newcommand{\kbh}{\hat{\bvec{k}}}
\newcommand{\nbh}{\hat{\bvec{n}}}
\newcommand{\pvbh}{\hat{\bvec{p}}}
\newcommand{\qbh}{\hat{\bvec{q}}}
\newcommand{\Xbh}{\hat{\bvec{X}}}
\newcommand{\rbh}{\hat{\bvec{r}}}
\newcommand{\xbh}{\hat{\bvec{x}}}
\newcommand{\ybh}{\hat{\bvec{y}}}
\newcommand{\zbh}{\hat{\bvec{z}}}
\newcommand{\betabh}{\hat{\bfbeta}}
\newcommand{\tbh}{\hat{\bfth}}
\newcommand{\pbh}{\hat{\bfvphi}}
\newcommand{\dt}{\Delta\tau}
\newcommand{\kmag}{|\bvec{k}|}
\newcommand{\pmag}{|\bvec{p}|}
\newcommand{\qmag}{|\bvec{q}|}
\newcommand{\oas}{\ensuremath{\mathcal{O}(\alpha_s)}}
\newcommand{\vtxb}{\ensuremath{\Lambda_\mu(p',p)}}
\newcommand{\vtxp}{\ensuremath{\Lambda^\mu(p',p)}}
\newcommand{\pwqp}{e^{i\bvec{q}\cdot\bvec{r}}}
\newcommand{\pwqm}{e^{-i\bvec{q}\cdot\bvec{r}}}
\newcommand{\gsa}[1]{\ensuremath{\bb#1\kb_0}}
\newcommand{\oer}[1]{\mathcal{O}\left(\frac{1}{\qmag^{#1}}\right)}
\newcommand{\nub}[1]{\overline{\nu^{#1}}}
\newcommand{\epf}{E_\bvec{p}}
\newcommand{\epfp}{E_{\bvec{p}'}}
\newcommand{\eka}{E_{\alpha\kappa}}
\newcommand{\ekaq}{(E_{\alpha\kappa})^2}
\newcommand{\ekap}{E_{\alpha'\kappa}}
\newcommand{\ekpa}{E+{\alpha\kappa_+}}
\newcommand{\ekma}{E_{\alpha\kappa_-}}
\newcommand{\ekp}{E_{\kappa_+}}
\newcommand{\ekm}{E_{\kappa_-}}
\newcommand{\ekpap}{E_{\alpha'\kappa_+}}
\newcommand{\ekmap}{E_{\alpha'\kappa_-}}
\newcommand{\yjm}[1]{\mathcal{Y}_{jm}^{#1}}
\newcommand{\ysa}[3]{\mathcal{Y}_{#1,#2}^{#3}}
\newcommand{\yss}[2]{\mathcal{Y}_{#1}^{#2}}
\newcommand{\Dj}{\ensuremath{\mathscr{D}}}
\newcommand{\ysc}{\tilde{y}}
\newcommand{\enm}{\varepsilon_{NM}}
\newcommand{\Scg}[6]
	{\ensuremath{S^{#1}_{#4}\:\vphantom{S}^{#2}_{#5}
 	 \:\vphantom{S}^{#3}_{#6}\,}}
\newcommand{\Kmat}[6]
	{\ensuremath{K\left[\begin{array}{ccc} 
	#1 & #2 & #3 \\ #4 & #5 & #6\end{array}\right]}}
\newcommand{\irt}{\ensuremath{\frac{1}{\sqrt{2}}}}
\newcommand{\sirt}{\ensuremath{\tfrac{1}{\sqrt{2}}}}
\newcommand{\irth}{\ensuremath{\frac{1}{\sqrt{3}}}}
\newcommand{\sirth}{\ensuremath{\tfrac{1}{\sqrt{3}}}}
\newcommand{\irs}{\ensuremath{\frac{1}{\sqrt{6}}}}
\newcommand{\sirs}{\ensuremath{\tfrac{1}{\sqrt{6}}}}
\newcommand{\tors}{\ensuremath{\frac{2}{\sqrt{6}}}}
\newcommand{\stors}{\ensuremath{\tfrac{2}{\sqrt{6}}}}
\newcommand{\rtoth}{\ensuremath{\sqrt{\frac{2}{3}}}}
\newcommand{\rthot}{\ensuremath{\frac{\sqrt{3}}{2}}}
\newcommand{\ithrt}{\ensuremath{\frac{1}{3\sqrt{2}}}}
\newcommand{\Tg}{\ensuremath{\mathsf{T}}}
\newcommand{\irrep}[1]{\ensuremath{\mathbf{#1}}}
\newcommand{\cirrep}[1]{\ensuremath{\overline{\mathbf{#1}}}}
\newcommand{\Fij}{\ensuremath{\hat{F}_{ij}}}
\newcommand{\Fqij}{\ensuremath{\hat{F}^{(qq)}_{ij}}}
\newcommand{\Fsij}{\ensuremath{\hat{F}^{(qs)}_{ij}}}
\newcommand{\Opij}{\mathcal{O}^p_{ij}}
\newcommand{\fpij}{f_p(r_{ij})}
\newcommand{\titj}{\bftau_i\cdot\bftau_j}
\newcommand{\sisj}{\bfsig_i\cdot\bfsig_j}
\newcommand{\Sij}{S_{ij}}
\newcommand{\LS}{\bvec{L}_{ij}\cdot\bvec{S}_{ij}}
\newcommand{\TT}{\Tg_i\cdot\Tg_j}
\newcommand{\chet}{\ensuremath{\chi ET}}
\newcommand{\chpt}{\ensuremath{\chi PT}}
\newcommand{\chsy}{\ensuremath{\chi\mbox{symm}}}
\newcommand{\lchi}{\ensuremath{\Lambda_\chi}}
\newcommand{\lcon}{\ensuremath{\Lambda_{QCD}}}
\newcommand{\dcpsi}{\ensuremath{\bar{\psi}}}
\newcommand{\dc}[1]{\ensuremath{\overline{#1}}}
\newcommand{\dcpsip}{\ensuremath{\bar{\psi}^{(+)}}}
\newcommand{\psip}{\ensuremath{{\psi}^{(+)}}}
\newcommand{\dcpsim}{\ensuremath{\bar{\psi}^{(-)}}}
\newcommand{\psim}{\ensuremath{{\psi}^{(-)}}}
\newcommand{\llo}{\ensuremath{\mathcal{L}^{(0)}_{\chet}}}
\newcommand{\lchet}{\ensuremath{\mathcal{L}_{\chi}}}
\newcommand{\hchet}{\ensuremath{\mathcal{H}_{\chi}}}
\newcommand{\Hd}{\ensuremath{\mathcal{H}}}
\newcommand{\Dmu}{\ensuremath{\mathcal{D}_\mu}}
\newcommand{\Dsl}{\ensuremath{\slashed{\mathcal{D}}}}
\newcommand{\comm}[2]{\ensuremath{\left[#1,#2\right]}}
\newcommand{\acomm}[2]{\ensuremath{\left\{#1,#2\right\}}}
\newcommand{\ev}[1]{\ensuremath{\bb\hat{#1}\kb}}
\newcommand{\exv}[1]{\ensuremath{\bb{#1}\kb}}
\newcommand{\evt}[1]{\ensuremath{\bb{#1}(\tau)\kb}}
\newcommand{\evm}[1]{\ensuremath{\bb{#1}\kb_M}}
\newcommand{\evv}[1]{\ensuremath{\bb{#1}\kb_V}}
\newcommand{\ovl}[2]{\ensuremath{\bb{#1}|{#2}\kb}}
\newcommand{\pd}{\partial}
\newcommand{\pnpd}[2]{\frac{\partial{#1}}{\partial{#2}}}
\newcommand{\pppd}[1]{\frac{\partial{\hphantom{#1}}}{\partial{#1}}}
\newcommand{\plmu}{\partial_\mu}
\newcommand{\plnu}{\partial_\nu}
\newcommand{\pumu}{\partial^\mu}
\newcommand{\punu}{\partial^\nu}
\newcommand{\mcdf}{\delta^{(4)}(p_f-p_i-q)}
\newcommand{\ecdf}{\delta(E_f-E_i-\nu)}
\newcommand{\tr}{\mbox{Tr }}
\newcommand{\lxr}{\ensuremath{SU(2)_L\times SU(2)_R}}
\newcommand{\gV}[2]{\ensuremath{(\gamma^{-1})^{#1}_{\hphantom{#1}{#2}}}}
\newcommand{\gVd}[2]{\ensuremath{\gamma^{#1}_{\hphantom{#1}{#2}}}}
\newcommand{\LpV}[1]{\ensuremath{\Lambda^{#1}V}}
\newcommand{\hatH}{\ensuremath{\hat{H}}}
\newcommand{\hath}{\ensuremath{\hat{h}}}
\newcommand{\eht}{\ensuremath{e^{-\tau\hat{H}}}}
\newcommand{\ehdt}{\ensuremath{e^{-\Delta\tau\hat{H}}}}
\newcommand{\ehtm}{\ensuremath{e^{-\tau(\hat{H}-E_V)}}}
\newcommand{\ehdtm}{\ensuremath{e^{-\Delta\tau(\hat{H}-E_V)}}}
\newcommand{\Oop}{\ensuremath{\mathcal{O}}}
\newcommand{\Gop}{\ensuremath{\hat{\mathcal{G}}}}
\newcommand{\SU}[1]{\ensuremath{SU({#1})}}
\newcommand{\U}[1]{\ensuremath{U({#1})}}
\newcommand{\proj}[1]{\ensuremath{\ket{#1}\bra{#1}}}
\newcommand{\su}[1]{\ensuremath{\mathfrak{su}({#1})}}
\newcommand{\ip}[2]{\ensuremath{\bvec{#1}\cdot\bvec{#2}}}
\newcommand{\norm}[1]{\ensuremath{\left| #1\right|^2}}
\newcommand{\rnorm}[1]{\ensuremath{\lvert #1\rvert}}
\newcommand{\pid}{\left(\begin{array}{cc} 1 & 0 \\ 0 & 1\end{array}\right)}
\newcommand{\psx}{\left(\begin{array}{cc} 0 & 1 \\ 1 & 0\end{array}\right)}
\newcommand{\psy}{\left(\begin{array}{cc} 0 & -i \\ i & 0\end{array}\right)}
\newcommand{\psz}{\left(\begin{array}{cc} 1 & 0 \\ 0 & -1\end{array}\right)}
\newcommand{\ua}{\uparrow}
\newcommand{\da}{\downarrow}
\newcommand{\deln}{\delta_{i_1 i_2\ldots i_n}}
\newcommand{\GabRR}{G_{\alpha\beta}(\bfR,\bfR')}
\newcommand{\GRR}{G(\bfR,\bfR')}
\newcommand{\GfRR}{G_0(\bfR,\bfR')}
\newcommand{\GRiR}{G(\bfR_i,\bfR_{i-1})}
\newcommand{\GRRs}[2]{G(\bfR_{#1},\bfR_{#2})}
\newcommand{\Gdgn}{\Gamma_{\Delta,\gamma N}}
\newcommand{\Gdgnb}{\overline\Gamma_{\Delta,\gamma N}}
\newcommand{\GJT}{\Gamma_{LS}^{JT}(k)}
\newcommand{\GJTa}[2]{\Gamma^{#1}_{#2}}
\newcommand{\GtwJTa}[2]{\tilde{\Gamma}_{#1}^{#2}}
\newcommand{\Gtw}{\tilde{\Gamma}}
\newcommand{\Gbar}{\overline{\Gamma}}
\newcommand{\Gtil}{\tilde{\Gamma}}
\newcommand{\Gpndb}{\overline{\Gamma}_{\pi N,\Delta}}
\newcommand{\GbNgn}{{\overline{\Gamma}}_{N^*,\gamma N}}
\newcommand{\GNgn}{\Gamma_{N^*,\gamma N}}
\newcommand{\GbNmb}{{\overline{\Gamma}}_{N^*,MB}}
\newcommand{\Lg}[2]{\ensuremath{L^{#1}_{\hphantom{#1}{#2}}}}
\newcommand{\psik}{\ensuremath{\left(\begin{matrix}\psi_1 \\ \psi_2\end{matrix}\right)}}
\newcommand{\psib}{\ensuremath{\left(\begin{matrix}\psi^*_1&\psi^*_2\end{matrix}\right)}}
\newcommand{\Gf}{\ensuremath{\frac{1}{E-H_0}}}
\newcommand{\Gv}{\ensuremath{\frac{1}{E-H_0-\vnres}}}
\newcommand{\Gx}{\ensuremath{\frac{1}{E-H_0-V}}}
\newcommand{\Gex}{\ensuremath{\mathcal{G}}}
\newcommand{\Gfpm}{\ensuremath{\frac{1}{E-H_0\pm i\epsilon}}}
\newcommand{\vres}{v_R}
\newcommand{\vnres}{v}
\newcommand{\tpz}{\ensuremath{^3P_0}}
\newcommand{\tres}{t_R}
\newcommand{\tsr}{t^R}
\newcommand{\tsnr}{t^{NR}}
\newcommand{\trest}{\tilde{t}_R}
\newcommand{\tnres}{t}
\newcommand{\Pt}{P_{12}}
\newcommand{\Sz}{\ket{S_0}}
\newcommand{\Sa}{\ket{S^{(-1)}_1}}
\newcommand{\Sb}{\ket{S^{(0)}_1}}
\newcommand{\Sc}{\ket{S^{(+1)}_1}}
\newcommand{\sbasis}{\ket{s_1 s_2; m_1 m_2}}
\newcommand{\Sbasis}{\ket{s_1 s_2; S M}}
\newcommand{\sket}[2]{\ket{{#1}\,{#2}}}
\newcommand{\sbra}[2]{\bra{{#1}\,{#2}}}
\newcommand{\psmket}{\ket{\bvec{p};s\,m}}
\newcommand{\cket}{\ket{\bvec{p};s_1 s_2\,m_1 m_2}}
\newcommand{\hket}{\ket{\bvec{p};s_1 s_2\,\lambda_1\lambda_2}}
\newcommand{\hkets}{\ket{s\,\lambda}}
\newcommand{\phkets}{\ket{\bvec{p};s\,\lambda}}
\newcommand{\klsjm}{\ket{p;\ell s; j m}}
\newcommand{\pq}{\bvec{p}_q}
\newcommand{\pqb}{\bvec{p}_{\qbar}}
\newcommand{\mps}[1]{\frac{d^3{#1}}{(2\pi)^{3/2}}}
\newcommand{\mpsf}[1]{\frac{d^3{#1}}{(2\pi)^{3}}}
\newcommand{\du}[1]{u_{\bvec{#1},s}}
\newcommand{\dv}[1]{v_{\bvec{#1},s}}
\newcommand{\cdu}[1]{\overline{u}_{\bvec{#1},s}}
\newcommand{\cdv}[1]{\overline{v}_{\bvec{#1},s}}
\newcommand{\dus}[2]{u_{\bvec{#1},{#2}}}
\newcommand{\dvs}[2]{v_{\bvec{#1},{#2}}}
\newcommand{\cdus}[2]{\overline{u}_{\bvec{#1},{#2}}}
\newcommand{\cdvs}[2]{\overline{v}_{\bvec{#1},{#2}}}
\newcommand{\bop}[1]{b_{\bvec{#1},s}}
\newcommand{\dop}[1]{d_{\bvec{#1},s}}
\newcommand{\bops}[2]{b_{\bvec{#1},{#2}}}
\newcommand{\dops}[2]{d_{\bvec{#1},{#2}}}
\newcommand{\mev}{\mbox{ MeV}}
\newcommand{\gev}{\mbox{ GeV}}
\newcommand{\fmi}{\mbox{ fm}}
\newcommand{\M}{\mathcal{M}}
\newcommand{\Smat}{\mathcal{S}}
\newcommand{\JLSTh}{JLST\lambda}
\newcommand{\Tpg}{T_{\pi N,\gamma N}}
\newcommand{\tpg}{t_{\pi N,\gamma N}}
\newcommand{\vmbmb}{\ensuremath{v_{M'B',MB}}}
\newcommand{\tmbgn}{\ensuremath{t_{MB,\gamma N}}}
\newcommand{\Tonon}{\ensuremath{T_{\omega N,\omega N}}}
\newcommand{\tonon}{\ensuremath{t_{\omega N,\omega N}}}
\newcommand{\tronon}{\ensuremath{t^R_{\omega N,\omega N}}}
\newcommand{\Tpnpn}{\ensuremath{T_{\pi N,\pi N}}}
\newcommand{\Tonpn}{\ensuremath{T_{\omega N,\pi N}}}
\newcommand{\tonpn}{\ensuremath{t_{\omega N,\pi N}}}
\newcommand{\tronpn}{\ensuremath{t^R_{\omega N,\pi N}}}
\newcommand{\Tongn}{\ensuremath{T_{\omega N,\gamma N}}}
\newcommand{\tongn}{\ensuremath{t_{\omega N,\gamma N}}}
\newcommand{\trongn}{\ensuremath{t^R_{\omega N,\gamma N}}}
\newcommand{\vmbgn}{\ensuremath{v_{MB,\gamma N}}}
\newcommand{\vpngn}{\ensuremath{v_{\pi N,\gamma N}}}
\newcommand{\vongn}{\ensuremath{v_{\omega N,\gamma N}}}
\newcommand{\vonpn}{\ensuremath{v_{\omega N,\pi N}}}
\newcommand{\vpnpn}{\ensuremath{v_{\pi N,\pi N}}}
\newcommand{\vonon}{\ensuremath{v_{\omega N,\omega N}}}
\newcommand{\vrngn}{\ensuremath{v_{\rho N,\gamma N}}}
\newcommand{\tjtmbmb}{\ensuremath{t^{JT}_{M'B',MB}}}
\newcommand{\tjlsmngn}{\ensuremath{t^{JT}_{L'S'M'N',\lag\lN T_{N,z}}}}
\newcommand{\tjlsmbgn}{\ensuremath{t^{JT}_{LSMB,\lag \lN T_{N,z}}}}
\newcommand{\vjlsmngn}{\ensuremath{v^{JT}_{L'S'M'N',\lag \lN T_{N,z}}}}
\newcommand{\vjlsmbgn}{\ensuremath{v^{JT}_{LSMB,\lag \lN T_{N,z}}}}
\newcommand{\tjlsmnmb}{\ensuremath{t^{JT}_{L'S'M'N',LSMB}}}
\newcommand{\Tjlsmbmb}{\ensuremath{T^{JT}_{LSMB,L'S'M'B'}}}
\newcommand{\tjlsmbmb}{\ensuremath{t^{JT}_{LSMB,L'S'M'B'}}}
\newcommand{\tjlsmnpn}{\ensuremath{t^{JT}_{L'S'M'N',\ell \pi N}}}
\newcommand{\tjlsmbpn}{\ensuremath{t^{JT}_{LSMB,\ell \pi N}}}
\newcommand{\vjlsmnpn}{\ensuremath{v^{JT}_{L'S'M'N',\ell \pi N}}}
\newcommand{\vjlsmnmb}{\ensuremath{v^{JT}_{L'S'M'N',LSMB}}}
\newcommand{\vjlsmbpn}{\ensuremath{v^{JT}_{LSMB,\ell \pi N}}}
\newcommand{\Tjlsmngn}{\ensuremath{t^{R,JT}_{L'S'M'N',\lag\lN T_{N,z}}}}
\newcommand{\Tjlsmbgn}{\ensuremath{t^{R,JT}_{LSMB,\lag \lN T_{N,z}}}}
\newcommand{\Tfjlsmbgn}{\ensuremath{T^{JT}_{LSMB,\lag \lN T_{N,z}}}}
\newcommand{\Tjlsmnmb}{\ensuremath{t^{R,JT}_{L'S'M'N',LSMB}}}
\newcommand{\Tjlsmnpn}{\ensuremath{t^{R,JT}_{L'S'M'N',\ell \pi N}}}
\newcommand{\Tjlsmbpn}{\ensuremath{t^{R,JT}_{LSMB,\ell \pi N}}}
\newcommand{\Gbjlsi}{\ensuremath{{\Gamma}^{JT}_{LSMB,N^*_i}}}
\newcommand{\Gbjlspi}{\ensuremath{{\Gamma}^{JT}_{L'S'M'B',N^*_i}}}
\newcommand{\Gjlsi}{\ensuremath{\overline{\Gamma}^{JT}_{LSMB,N^*_i}}}
\newcommand{\Gijls}{\ensuremath{\overline{\Gamma}^{JT}_{N^*_i,LSMB}}}
\newcommand{\Gbijls}{\ensuremath{{\Gamma}^{JT}_{N^*_i,LSMB}}}
\newcommand{\Gjpn}{\ensuremath{\overline{\Gamma}^{JT}_{N^*_j,\ell\pn}}}
\newcommand{\Gign}{\ensuremath{\overline{\Gamma}^{JT}_{N^*_i,\lag\lN T_{N,z}}}}
\newcommand{\Gbign}{\ensuremath{{\Gamma}^{JT}_{N^*_i,\lag\lN T_{N,z}}}}
\newcommand{\Gjlsj}{\ensuremath{\overline{\Gamma}^{JT}_{LSMB,N^*_j}}}
\newcommand{\Gjem}{\ensuremath{\overline{\Gamma}^{JT}_{N^*_j,\lag\lN T_{N,z}}}}
\newcommand{\Ljtlsmbn}{\ensuremath{\Lambda^{JT}_{N^*LSMB}}}
\newcommand{\Drij}{\ensuremath{\mathcal{D}^{-1}_{ij}}}
\newcommand{\Mbres}{\ensuremath{M^{(0)}_{N^*}}}
\newcommand{\Cjtnlsmb}{\ensuremath{C^{JT}_{N^*LSMB}}}
\newcommand{\Ljtnlsmb}{\ensuremath{\Lambda^{JT}_{N^*LSMB}}}
\newcommand{\knstar}{\ensuremath{k_{N^*}}}
\newcommand{\vonen}{\ensuremath{v_{\omega N,\eta N}}}
\newcommand{\vonpd}{\ensuremath{v_{\omega N,\pi\Delta}}}
\newcommand{\vonsn}{\ensuremath{v_{\omega N,\sigma N}}}
\newcommand{\vonrn}{\ensuremath{v_{\omega N,\rho N}}}
\newcommand{\gnon}{\ensuremath{\gamma N\to \omega N}}
\newcommand{\gnpn}{\ensuremath{\gamma N\to \pi N}}
\newcommand{\gnen}{\ensuremath{\gamma N\to \eta N}}
\newcommand{\gpop}{\ensuremath{\gamma p\to \omega p}}
\newcommand{\gpep}{\ensuremath{\gamma p\to \eta p}}
\newcommand{\gnten}{\ensuremath{\gamma n\to \eta n}}
\newcommand{\gppzp}{\ensuremath{\gamma p\to \pi^0 p}}
\newcommand{\gpppn}{\ensuremath{\gamma p\to \pi^+ n}}
\newcommand{\gnpmp}{\ensuremath{\gamma n\to \pi^- p}}
\newcommand{\gnpzn}{\ensuremath{\gamma n\to \pi^0 n}}
\newcommand{\gppzep}{\ensuremath{\gamma p\to \pi^0 \eta p}}
\newcommand{\pnen}{\ensuremath{\pi N\to \eta N}}
\newcommand{\pnon}{\ensuremath{\pi N\to \omega N}}
\newcommand{\pnmb}{\ensuremath{\pi N\to MB}}
\newcommand{\gnmb}{\ensuremath{\gamma N\to M\!B}}
\newcommand{\onon}{\ensuremath{\omega N\to \omega N}}
\newcommand{\pmpon}{\ensuremath{\pi^- p\to \omega n}}
\newcommand{\pnpn}{\ensuremath{\pi N\to \pi N}}
\newcommand{\Gon}{\ensuremath{G_{0,\omega N}}}
\newcommand{\Gpn}{\ensuremath{G_{0,\pi N}}}
\newcommand{\rhomb}{\ensuremath{\rho_{MB}}}
\newcommand{\rhoon}{\ensuremath{\rho_{\omega N}}}
\newcommand{\rhopn}{\ensuremath{\rho_{\pi N}}}
\newcommand{\kon}{\ensuremath{k_{\omega N}}}
\newcommand{\kpn}{\ensuremath{k_{\pi N}}}
\newcommand{\Gmb}{\ensuremath{G_{0,MB}}}
\newcommand{\Tmbgn}{\ensuremath{T_{MB,\gamma N}}}
\newcommand{\vmbpgn}{\ensuremath{v_{M'B',\gamma N}}}
\newcommand{\pntpn}{\ensuremath{\pi N\!\to\!\pi N}}
\newcommand{\pnten}{\ensuremath{\pi N\!\to\!\eta N}}
\newcommand{\pnton}{\ensuremath{\pi N\!\to\!\omega N}}
\newcommand{\epos}{\ensuremath{\slashed{\epsilon}_{\lambda_\omega}}}
\newcommand{\epo}{\ensuremath{{\epsilon}_{\lambda_\omega}}}
\newcommand{\elevi}{\ensuremath{{\epsilon}_{\alpha\beta\gamma\delta}}}
\newcommand{\eps}{\ensuremath{\epsilon}}
\newcommand{\krho}{\ensuremath{\kappa_\rho}}
\newcommand{\komg}{\ensuremath{\kappa_\omega}}
\newcommand{\komga}{\ensuremath{\kappa^t_\omega}}
\newcommand{\doh}{\ensuremath{d^{(\half)}_{\lambda'\lambda}}}
\newcommand{\dohm}{\ensuremath{d^{(\half)}_{-\lambda,-\lambda'}}}
\newcommand{\dohmo}{\ensuremath{d^{(\half)}_{\lambda',-\half}}}
\newcommand{\dohpo}{\ensuremath{d^{(\half)}_{\lambda',+\half}}}
\newcommand{\Lor}[2]{\ensuremath{\Lambda^{#1}_{\hphantom{#1}{#2}}}}
\newcommand{\ILor}[2]{\ensuremath{\Lambda_{#1}^{\hphantom{#1}{#2}}}}
\newcommand{\LorT}[2]{\ensuremath{[\Lambda^T]^{#1}_{\hphantom{#1}{#2}}}}
\newcommand{\dsdo}{{\frac{d\sigma}{d\Omega}}}
\newcommand{\dspdo}{\ensuremath{{\frac{d\sigma_\pi}{d\Omega}}}}
\newcommand{\dsgdo}{\ensuremath{{\frac{d\sigma_\gamma}{d\Omega}}}}
\newcommand{\chipd}{\ensuremath{\chi^2/N_d}}
\newcommand{\chipda}{\ensuremath{\chi^2(\alpha)/N_d}}
\newcommand{\bpop}{\ensuremath{\bvec{p}'_1}}
\newcommand{\bptp}{\ensuremath{\bvec{p}'_2}}
\newcommand{\bpip}{\ensuremath{\bvec{p}'_i}}
\newcommand{\bpo}{\ensuremath{\bvec{p}_1}}
\newcommand{\bpt}{\ensuremath{\bvec{p}_2}}
\newcommand{\bpi}{\ensuremath{\bvec{p}_i}}
\newcommand{\bqo}{\ensuremath{\bvec{q}_1}}
\newcommand{\bqt}{\ensuremath{\bvec{q}_2}}
\newcommand{\bqi}{\ensuremath{\bvec{q}_i}}
\newcommand{\bQ}{\ensuremath{\bvec{Q}}}
\newcommand{\bq}{\ensuremath{\bvec{q}}}
\newcommand{\ketq}{\ensuremath{\ket{\bqo,\bqt}}}
\newcommand{\ketqc}{\ensuremath{\ket{\bQ,\bq}}}
\newcommand{\bP}{\ensuremath{\bvec{P}}}
\newcommand{\bPp}{\ensuremath{\bvec{P}'}}
\newcommand{\bpr}{\ensuremath{\bvec{p}}}
\newcommand{\bprp}{\ensuremath{\bvec{p}'}}
\newcommand{\ketPsiq}{\ensuremath{\ket{\Psi_{\bq}^{(\pm)}}}}
\newcommand{\ketPsiqQ}{\ensuremath{\ket{\Psi_{\bQ,\bq}^{(\pm)}}}}
\newcommand{\Ld}{\ensuremath{\mathcal{L}}}
\newcommand{\ps}{\mbox{ps}}
\newcommand{\fndp}{f_{N\Delta\pi}}
\newcommand{\fndr}{f_{N\Delta\rho}}
\newcommand{\said}{{\sc said}}
\newcommand{\ret}{\ensuremath{\langle{\tt ret}\rangle}}
\newcommand{\ddf}[1]{\ensuremath{\delta^{(#1)}}}
\newcommand{\Tpp}{\ensuremath{T_{\pi\pi}}}
\newcommand{\Kpp}{\ensuremath{K_{\pi\pi}}}
\newcommand{\Tpe}{\ensuremath{T_{\pi\eta}}}
\newcommand{\Kpe}{\ensuremath{K_{\pi\eta}}}
\newcommand{\Tep}{\ensuremath{T_{\eta\pi}}}
\newcommand{\Kep}{\ensuremath{K_{\eta\pi}}}
\newcommand{\Tee}{\ensuremath{T_{\eta\eta}}}
\newcommand{\Kee}{\ensuremath{K_{\eta\eta}}}
\newcommand{\Tpig}{\ensuremath{T_{\pi\gamma}}}
\newcommand{\Kpig}{\ensuremath{K_{\pi\gamma}}}
\newcommand{\oKpig}{\ensuremath{\overline{K}_{\pi\gamma}}}
\newcommand{\tKpig}{\ensuremath{\tilde{K}_{\pi\gamma}}}
\newcommand{\Teg}{\ensuremath{T_{\eta\gamma}}}
\newcommand{\Keg}{\ensuremath{K_{\eta\gamma}}}
\newcommand{\Kab}{\ensuremath{K_{\alpha\beta}}}
\newcommand{\R}{\ensuremath{\mathbb{R}}}
\newcommand{\C}{\ensuremath{\mathbb{C}}}
\newcommand{\Ezp}{\ensuremath{E^{\pi}_{0+}}}
\newcommand{\Eze}{\ensuremath{E^{\eta}_{0+}}}
\newcommand{\Ga}{\ensuremath{\Gamma_\alpha}}
\newcommand{\Gb}{\ensuremath{\Gamma_\beta}}
\newcommand{\RH}{\ensuremath{\mathcal{R}\!\!-\!\!\mathcal{H}}}
\newcommand{\calT}{\mathcal{T}}
\newcommand{\maid}{{\sc maid}}
\newcommand{\Kbar}{\ensuremath{\overline{K}}}
\newcommand{\zbar}{\ensuremath{\overline{z}}}
\newcommand{\kbar}{\ensuremath{\overline{k}}}
\newcommand{\dom}{\ensuremath{\mathcal{D}}}
\newcommand{\domi}[1]{\ensuremath{\mathcal{D}_{#1}}}
\newcommand{\pbar}{\ensuremath{\overline{p}}}
\newcommand{\Nab}{\ensuremath{N_{\alpha\beta}}}
\newcommand{\Nee}{\ensuremath{N_{\eta\eta}}}
\newcommand{\dth}[1]{\delta^{(3)}(#1)}
\newcommand{\dfo}[1]{\delta^{(4)}(#1)}
\newcommand{\intk}{\int\!\!\frac{d^3\! k}{(2\pi)^3}}
\newcommand{\intkg}{\int\!\!{d^3\! k_\gamma}}
\newcommand{\intks}{\int\!\!{d^3\! k_\sigma}}
\newcommand{\nch}{\ensuremath{N_{\mbox{ch}}}}
\newcommand{\nc}{\ensuremath{N_{ch}}}
\newcommand{\re}{\ensuremath{\mbox{Re }\!}}
\newcommand{\im}{\ensuremath{\mbox{Im }\!}}
\newcommand{\EetaS}{\ensuremath{E^\eta_{0+}}}
\newcommand{\EpiS}{\ensuremath{E^\pi_{0+}}}
\newcommand{\tobull}{\ensuremath{\to}}
\newcommand{\Kcm}{\ensuremath{K_{CM}}}

\newcommand{\gn}{\ensuremath{\gamma N}}
\newcommand{\gp}{\ensuremath{\gamma p}}
\newcommand{\geta}{\ensuremath{\gamma \eta}}
\newcommand{\pp}{\ensuremath{pp}}
\newcommand{\pn}{\ensuremath{\pi N}}
\newcommand{\phn}{\ensuremath{\pi d}}
\newcommand{\en}{\ensuremath{\eta N}}
\newcommand{\epn}{\ensuremath{\eta' N}}
\newcommand{\pD}{\ensuremath{\pi \Delta}}
\newcommand{\sn}{\ensuremath{\sigma N}}
\newcommand{\rn}{\ensuremath{\rho N}}
\newcommand{\on}{\ensuremath{\omega N}}
\newcommand{\ppn}{\ensuremath{\pi\pi N}}
\newcommand{\kn}{\ensuremath{KN}}
\newcommand{\ky}{\ensuremath{KY}}
\newcommand{\kl}{\ensuremath{K\Lambda}}
\newcommand{\ks}{\ensuremath{K\Sigma}}
\newcommand{\bn}{\ensuremath{eN}}
\newcommand{\bpn}{\ensuremath{e\pi N}}
\newcommand{\fpo}{\ensuremath{5\oplus 1}}
\newcommand{\faoe}{{\sc FA08}}
\newcommand{\fpoe}{{\sc FP08}}
\newcommand{\fsoe}{{\sc FS08}}
\newcommand{\psic}{\ensuremath{\psi_{n\kappa jm}}}

\newcommand{\itPFP}{\textit{Physics for Future Presidents}}
\newcommand{\itaPFP}{\textit{PFP}}

\title{Comparing partial-wave amplitude
parametrization with dynamical models of meson-nucleon scattering}
\author{Mark W.\ Paris and Ron L.\ Workman}
\affiliation{
Data Analysis Center at the Center for Nuclear Studies,\\
Department of Physics\\
The George Washington University,
Washington, D.C. 20052}

\date{\today}
 
\begin{abstract}
Relationships between partial-wave amplitude parametrizations, in
particular the Chew-Mandelstam approach, and dynamical coupled-channel
models are established and investigated. A bare pole corresponding to
the $\Delta (1232)$ resonance, found in a recent dynamical-model fit
to $\pi-$ and $\omega-$meson production reactions, compares closely to
one found in a unitary multichannel partial-wave amplitude
parametrization of SAID. The model dependence of the bare pole
precludes a direct connection between the approaches but is suggestive
that the dynamical description and the phenomenological
parametrization are closely related.
\end{abstract}


\maketitle

\section{Introduction}
\label{sec:intro}
In this study, we outline both qualitative and 
quantitative relationships between dynamical models and 
the SAID approach to fitting meson production reactions. 
The SAID parametrization is based on a Chew-Mandelstam (CM)
approach\cite{Babelon:1976kv,Basdevant:1978tx} that has been
extensively applied in multichannel descriptions of hadronically and
electromagnetically induced reactions on the
proton\cite{Arndt:1985vj,Arndt:1989ww,Arndt:1995bj,Arndt:2001si,
Arndt:2003if,Arndt:2006bf,Prakhov:2005qb}.  We compare this with
recent multichannel dynamical model descriptions that assume a set of
well established $N$ and $\Delta$ resonances.

Meson scattering and production reactions (collectively,
``reactions'') account almost entirely for information available on
the resonance structure of the nucleon.  The resonances of the nucleon
encode a wealth of information on the non-perturbative regime of
quantum chromodynamics, the fundamental non-Abelian quantum field
theory of quarks and gluons, that is responsible for nuclear forces
and the interactions of effective hadronic degrees-of-freedom. Here we
elaborate on the connections between well-known and widely used
multichannel parametrization approaches, and dynamical model
approaches, which both satisfy unitarity at the two-body level.

An issue of recent debate\cite{Doring:2009bi} has been the interplay
of singularities arising from the iteration of colloquially termed
``non-pole'' or ``nonresonant'' interactions and those explicitly
added as bare states.  Within the CM approach, one may
ask\cite{Svarc:2008pc} whether the inclusion of poles in the
Chew-Mandelstam $K$ matrix is required. The relation between CM and
Heitler $K$-matrix\cite{Heitler:1941a} poles has also been
studied\cite{Workman:2008iv}.

The present SAID fits to pion-nucleon scattering and eta-nucleon
production data are based on multichannel
amplitudes\cite{Arndt:2006bf} for which, with the single exception of
the $P_{33}$ partial wave, the CM $K$ matrix, $\Kbar$ has been
constructed as a low-order polynomial in the center-of-mass complex
scattering energy, $E$, without poles. This construction may, however,
yield a pole in the Heitler $K$ matrix\cite{Paris:2010tz}, given the
relationship between $K$, the Heitler $K$ matrix, and $\Kbar$, the CM
$K$ matrix:
\begin{align}
\label{eqn:HKCMK}
K(E) &= \Kbar(E) \frac{1}{1-\re C(E) \Kbar(E)},
\end{align}
where $C(E)$ is termed the Chew-Mandelstam
`function,'\cite{Basdevant:1978tx} a diagonal matrix
in the space of included channels. Evidently, poles may be located at
real values of the scattering energy, $W=\re E$ when
$\det[1-\re C\Kbar]=0$\cite{Paris:2010tz}. A strong feature of the CM
approach then is that the unstable `particles' (identified with the 
baryon resonances) of the theory arise dynamically, in a sense, from
the proper analytic form of the CM parametrization in the process of
fitting to the observed data in the physical region, $W=\re E>0$. The
unstable particles are identified as poles of the $T$ matrix near the
physical region.

One may still include explicit poles in $\Kbar$ as is done in the
$P_{33}$ partial wave of Ref.\cite{Arndt:2006bf}. If poles are present
in $\Kbar$, how should they be interpreted? A more involved
question might ask for the effect of adding explicit CM $K$-matrix
poles in partial waves reproducing data {\it without} their inclusion.
Here, we restrict our discussion to the elastic $P_{33}$ partial wave
in the $\Delta (1232)$ resonance region. This partial wave has both
explicit CM and Heitler $K$-matrix poles and we compare them to
structures seen in a particular dynamical model.

In Section \ref{sec:mpa} we discuss general features of
each approach, while details are referred to in the
literature. Section \ref{sec:conc} contains a discussion of the 
results, possible extensions to this work, and conclusions.

\section{Model and parametrization approaches}
\label{sec:mpa}
This study is motivated in part by the observation that bare resonance
parameters of a recent SAID parametrization \cite{Arndt:2006bf} 
compare closely with those of recent dynamical model
calculations\cite{JuliaDiaz:2007kz,Paris:2008ig}. The
first subsection deals with details of the dynamical model and the
Chew-Mandelstam parametrization approaches relevant to understanding
this comparison.  The second subsection gives a more general
discussion of dynamics in the model and parametrization approaches.

\subsection{Dynamical model}
\label{subsec:respar}
We limit our discussion of the dynamical coupled-channel approaches to
the general features relevant to describing bare resonance parameters.
In this work, we study these model-dependent quantities and their
relation to physical, dressed resonance poles in the scattering, $S$
or transition, $T$ matrices. Details of the dynamical model can be
found elsewhere\cite{JuliaDiaz:2007kz,Paris:2008ig}.

The reaction observables are determined by the matrix elements of the
transition operator, $T$, a matrix in momentum, spin, and channel
spaces. The dynamical approach makes a model-dependent 
separation into `nonresonant' and `resonant' contributions:
\begin{align}
\label{eqn:Tconts}
T(E) &= t(E) + t^R(E).
\end{align}
Since the total $T$ matrix satisfies the (relativistic)
Lippmann-Schwinger equation
\begin{align}
\label{eqn:LS}
T(E) &= V + V G_0(E) T(E)
\end{align}
where $V$ is the (relativistic) tree-level interaction kernel, $G_0 =
[E - H_0]^{-1}$ is the many-body free particle propagator for the free
particle Hamiltonian, $H_0$ with physical masses, and $E\in\mathbb{C}$
is the complex scattering energy, the nonresonant and resonant
contributions satisfy
\begin{align}
\label{eqn:tnr}
t(E) &= v + v G_0(E) t(E), \\
\label{eqn:tres}
t^R(E) &= \Gbar^\dag(E) \frac{1}{E-H_0-\Sigma(E)}\Gbar(E),
\end{align}
respectively. Here, $v$ is the
non-resonant contribution to $V$, $\Gbar(\Gbar^\dagger)$ is the $MB\to
B'$($B\to MB'$) dressed vertex, $M$ and $B$ are meson and baryon
fields, $\Sigma(E)=\Gamma G_0(E) \Gbar^\dag(E)$ is the baryon
self-energy, and $\Gamma$ is the bare vertex. (Here, the $\dag$ symbol
is not Hermitian conjugation. It simply distinguishes $MB\to B'$ from
the inverse process.) The dressed vertex is related to the bare vertex
as $\Gbar=\Gamma(1+G_0 t(E))$. The resonances of the dynamical model
are determined by locating the set of complex energies
$\{E_r\}_{r=1}^{N_R}$ where
\begin{align}
\det [E_r - (M^{(0)} + \Sigma(E_r))] &= 0,
\end{align}
corresponding to poles of the $T$ (or $S$) matrix. Here $N_R$ is the
number of energies in a given partial wave for which the above
equation is satisfied. We note in passing that the number of resonance
poles need not be equal to the number of bare states with real
energies $M^{(0)}_i$\cite{Suzuki:2009nj}, where $i$ indexes one
specific, of a number of assumed, bare one-particle intermediate
states. We note that for the $P_{33}$ partial wave the poles of $T$
and the poles of $t^R$ are identical since there are no poles in the
non-resonant part $t$.  The pole mass, $M_r$ and total width,
$\Gamma_r$ of the baryon resonance are related to the pole position,
$E_r$ as
\begin{align}
M_r &= \re E_r, \\
-\shalf\Gamma_r &= \im E_r,
\end{align}
which are complicated functions of the bare parameters of the
Lagrangian. The bare mass, $M^{(0)}_i$ is dressed by interaction with
the meson and baryon fields of the Lagrangian.

It is appropriate to reiterate caveats associated with the bare
parameters of reaction theories in general and the particular
dynamical coupled-channel approach described here.  There are several
features of the dynamical approach that are model dependent and for
which extensive studies to quantify the precision of model predictions
are lacking. For example, there are no quantitative studies, to our
knowledge, that specify the accuracy of the bare parameters of the
Lagrangian within a given model approach, including the bare resonance
mass and width. The origin of the model dependence in the dynamical
approach stems, in part, from the application of the calculational
device of decomposing the $T$ matrix in Eq.\eqref{eqn:Tconts} into two
terms called, colloquially, nonresonant and resonant. The model
dependence of this decomposition may be viewed as a consequence of the
field redefinition ambiguities of the effective quantum field theory
that describes the hadronic degrees-of-freedom.  As shown in
Ref.\cite{Gegelia:2009py}, a field redefinition may be applied, for
example, to transform a heavy fermion field operator.  This has the
effect of shifting strength between nonresonant and resonant
mechanisms while leaving invariant the observables of the theory. The
resonance parameters, however, depend on the field redefinition and
therefore cannot be observables. We also reiterate the fact that the
CM parametrization approach makes no such decomposition of the $T$
matrix into nonresonant or resonant contributions. Therefore, it is
free of model dependencies that arise in such a decomposition.

We have, nevertheless, noticed a close numerical match between the bare
value of the bare pole position in the dynamical model and explicitly
included CM $K$ matrix pole in the $P_{33}$ partial wave of the
parametrization employed in Ref.\cite{Arndt:2006bf}. In the next
subsection we analyze this result in terms of the structure of
the CM parametrization form. Our hope is that by analyzing the
structure of both the model and parametrization forms and studying
their relationships that we may learn more about the nature of the
dynamics included within each approach.

\subsection{Chew-Mandelstam approach}
\label{subsec:CM}
We give a brief description of the CM parametrization form here. More
complete discussions of the form and its application to the
determination of the partial wave amplitudes in the multichannel
hadro- and photoproduction observed data are described in the
literature\cite{Babelon:1976kv,Basdevant:1978tx,Paris:2010tz}

The unitarity of the $S$ matrix implies a constraint, called
``unitarity,'' on the $T$ matrix, which may be concisely expressed as:
\begin{align}
\im T^{-1}(E) &=-\rho(E),
\end{align}
where $\rho =-\pi\delta(E-H_0)$ and the relationship between $T$ and
$S$ is given by $S=1+2i\rho T$.
This implies that the full $T$ matrix satisfies the Heitler equation:
\begin{align}
\label{eqn:Heitler}
T(E) &= K(E) + i K(E) \rho(E) T(E),
\end{align}
where $K^{-1} \equiv \re T^{-1}$. The Heitler $K$ matrix, $K$, has the
features of being a real-symmetric matrix in channel space and of
being free of threshold branch point singularities\cite{Zimmerman:1961aa}.
The Heitler $K$ matrix expressed in terms of the CM $K$ matrix,
$\Kbar$ is
\begin{align}
\label{eqn:KinvReln}
K^{-1}(E) &= \Kbar^{-1}(E) - \re C(E),
\end{align}
where $\re C$ is the Hilbert transform of $\rho$ (with one
subtraction). Solving for $K$ in the above relation yields
Eq.\eqref{eqn:HKCMK}. Parametrization of the matrix elements of
$\Kbar$ as functions
analytic in the finite complex $E$-plane yield, in general, matrix
elements of $K$ that are meromorphic in the energy. The $K$-matrix
poles that arise in this manner are sometimes related to the
poles of the $T$ matrix. This connection has been explored in
Ref.\cite{Workman:2008iv}. The $T$ matrix may then be expressed as
\begin{align}
\label{eqn:TCM}
T(E) &= \Kbar(E) + \Kbar(E) C(E) T(E)
\end{align}
directly in terms of the CM $K$ matrix.

Having established the representation of the $T$ matrix
in terms of the parametrized function $\Kbar$ we consider the form of
its matrix elements which include a single explicit pole. The $\Kbar$
matrix is then written as
\begin{align}
\label{eqn:KCMme}
\Kbar &= \frac{\gamma}{W-W_p} + \beta(W),
\end{align}
where $\gamma$ is a constant matrix in channel space and $\beta(W)$ is
an analytic matrix function of $W$, typically a polynomial.
This is, in fact, the form of the CM $K$ matrix used
in Ref.\cite{Arndt:2006bf} for the $P_{33}$ partial wave. 
Our fit to the complete \pn\ elastic and
$\pn\to\en$ reaction data gives, for the pole position of the CM $K$
matrix,
\begin{align}
\label{eqn:KCMpoleSAID}
W_p(P_{33}) &= 1381 \mev.
\end{align}
The position of the Heitler $K$-matrix pole in the SAID $P_{33}$
parametrization is related the CM $K$ matrix pole as:
\begin{align}
\label{eqn:KpoleSAID}
W_p^H(P_{33}) = W_p(P_{33}) 
  + \frac{\gamma_{\pn,\pn} \re C_{\pn}}{1-\beta_{\pn,\pn}\re C_{\pn}}
\end{align}
where, $C_{\pn}$ is the elastic, $\pn$ matrix element of the CM
function. The position of the Heitler $K$-matrix pole
is\cite{Workman:2008iv}
\begin{align}
\label{eqn:Kpoleval}
W_p^H(P_{33}) &= 1232 \mev
\end{align}
and is essentially model-independent\cite{Davidson:1990yk}.
Energy-dependent quantities in Eq.\eqref{eqn:KpoleSAID} are evaluated
at the Heitler $K$ matrix pole.

The fact that both the Heitler $K$ matrix, $K$, and the CM $K$ matrix,
$\Kbar$, have poles on the physical region at $W^H_p(P_{33})=1232$ MeV
and $W_p(P_{33})=1381$ MeV, respectively, constrains the
real part of $T_{\pn,\pn}(W^H_p(P_{33}))$ and the Chew-Mandelstam
function, $C_{\pn}(W_p(P_{33}))$. Employing Eqs.\eqref{eqn:Heitler}
and \eqref{eqn:TCM} for
energies below the point at which inelastic channels become important,
$W \lesssim 1450$ MeV (see Fig.\eqref{fig:invK}, below)
and taking the imaginary part gives the following expression:
\begin{align}
\label{eqn:imHK}
&K_{\pn,\pn}(W)\, \re T_{\pn,\pn}(W) = \Kbar_{\pn,\pn}(W) \nonumber \\
&\times\left[\frac{\re C_{\pn}(W)}{\rho_{\pn}(W)} 
       \im T_{\pn,\pn}(W) + \re T_{\pn,\pn}(W)\right].
\end{align}
Note that this relation holds in the elastic region only.
This relation provides constraints at the pole positions
$W^H_p$ and $W_p$ (where the $\pn$ partial wave designation, $P_{33}$
is to be understood from here forward):
\begin{align}
\label{eqn:Hpolreln}
\re T_{\pn,\pn}(W^H_p) &= 0 \\
\label{eqn:CMpolreln}
\tan \delta_{\pi}(W_p) + \frac{\rho_{\pn}(W_p)}{\re C_{\pn}(W_p)} &= 0,
\end{align}
which are satisfied at the values given in Eqs.\eqref{eqn:KCMpoleSAID} 
and \eqref{eqn:Kpoleval}, respectively. Here, $\tan\delta_\pi = \im \Tpnpn /
\re\Tpnpn$ is determined by the elastic $\pn\to\pn$ $P_{33}$ phase
shift.  Variability in the pole position is therefore directly linked
to our determination of $\re C_{\pn}(W)$ and, conversely, the
determination of $C_{\pn}(W_p)$ must satisfy the constraint of
Eq.\eqref{eqn:CMpolreln}; it is not model dependent. As the SAID
parametrization of $\re C_{\pn}(W)$ is fixed, apart from a subtraction
determining its zero point, and this value is not searched, the pole
position has remained essentially constant for different SAID
solutions even as new data and dispersion-relation constraints were
added. We also note that the $T$-matrix values at the Heitler
$K$-matrix pole, $T_{\pn,\pn}(W^H_p)=i$, and at the CM $K$-matrix
pole, $T_{\pn,\pn}(W_p) = -\rho_{\pn}(W_p) / C_{\pn}(W_p)$, are
independent of the $K$-matrix parametrization. A note of caution may
be appropriate here: Eq.\eqref{eqn:Hpolreln} appears similar to the
condition for locating the real part of the pole position of a
Breit-Wigner parametrization. We hope, however, that the preceding
discussion makes clear that the position of the Heitler $K$ matrix
pole bares no simple relation to model dependent and generally
non-unitary Breit-Wigner parametrization.

\begin{figure}
\includegraphics[width=8.5cm]{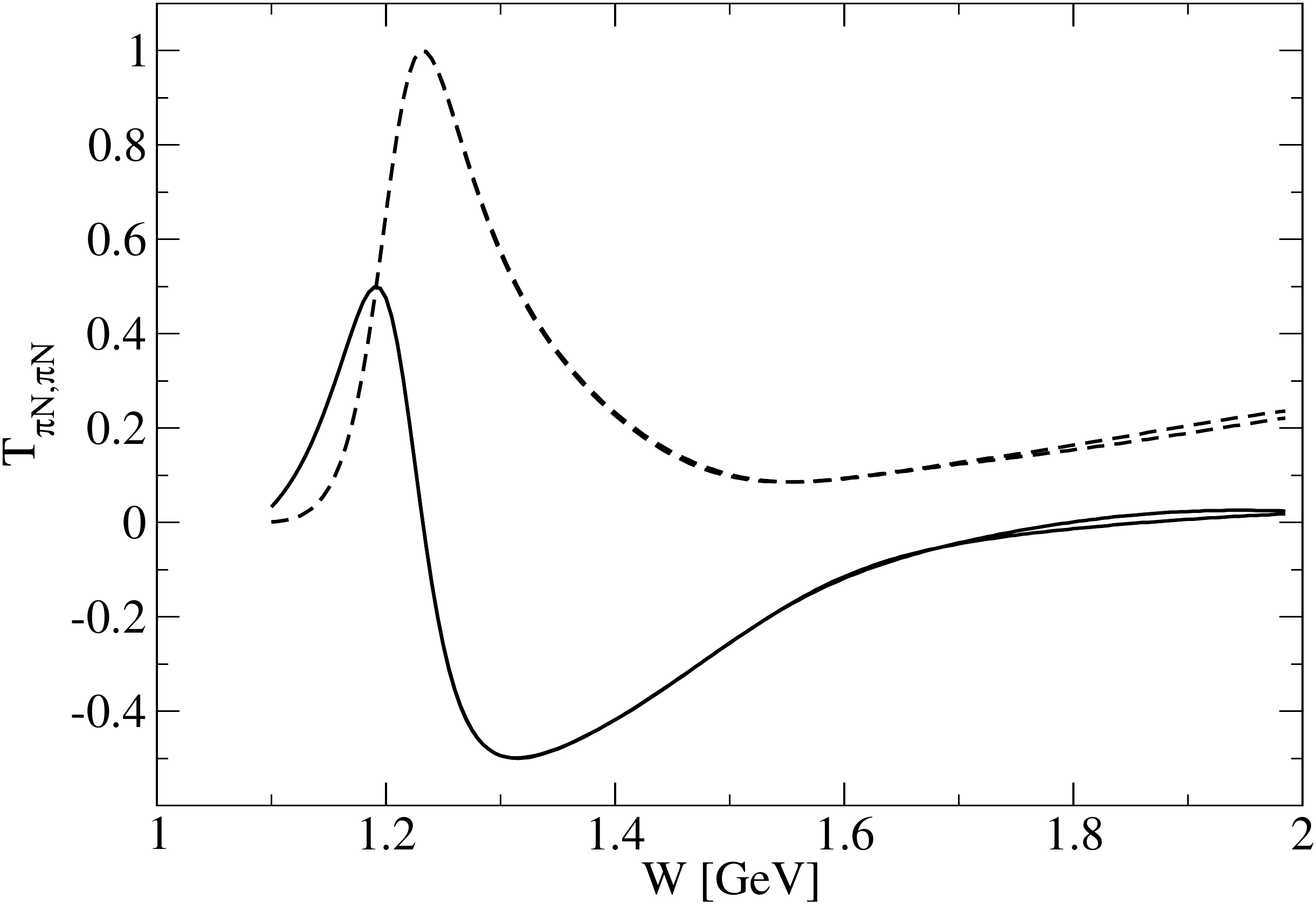}
\caption{\label{fig:pntm}Comparison of the {\sc SP06} $P_{33}$ partial
wave elastic $T$ matrix element\cite{Arndt:2006bf} (dimensionless)
having a sum of pole and non-pole $K$-matrix elements, and a fit using
the product $S$-matrix form, {\sc FP10}. The plotted real (solid) and
imaginary (dashed) parts are virtually identical over the resonance
region, producing equivalent fits to data.}
\end{figure}

If $\beta_{\pn, \pn}$ is set to zero and the $T$-matrix pole position
recalculated, its value shifts to $1297-i24$ MeV. Clearly the pole and
non-pole terms at the CM $K$-matrix level do not translate directly
into resonant and nonresonant contributions to the $T$ matrix.  This
separation is possible, phenomenologically, if one instead
parametrizes the $S$ matrix as $S_B S_R S_B$ with the resonance $T$
matrix, giving $S_R$, fitted as above but with $\beta_{\pn,\pn}$ set
to zero.  The $S_B$ is then similarly constructed with
$\gamma_{\pn,\pn}$ set to zero. A fit to data using this form,
{\sc FP10},
has been completed\cite{Workman:2010ip} and is plotted against the
standard form ({\sc SP06}) in Fig.\eqref{fig:pntm}.  The resulting Heitler
$K$-matrix and $T$-matrix poles are found at 1232 MeV and $1211-i50$
MeV respectively, as in the original fit. The CM $K$-matrix pole
shifts, in this case, to 1480 MeV.  Here, however, setting the
``background'' to zero (\textit{ie.}, $S_B\to 1$) has no effect on the
$T$-matrix pole position.

At higher energies, many resonances can be discovered by searching
for poles in the complex energy plane, without the explicit 
introduction of CM $K$-matrix poles. It would be interesting
to consider the effect of adding explicit poles to resonant partial
waves of this type. The resulting interplay of pole and non-pole
contributions could add new structures or replace resonances,
generated by non-pole terms, with pole-generated structures. 

\subsection{Comparison of dynamical model and CM parametrization}
\label{subsec:mpc}

Turning to the comparison of the dynamical model with the SAID CM
parametrization, Refs. \cite{Paris:2008ig} and \cite{Suzuki:2009nj}
give the values of bare resonance parameters in their Tables VI and
II, respectively. The value for the $P_{33}$ resonance is given as
\begin{align}
\label{eqn:barepole}
M^{(0)}(P_{33}) &= 1391 \mev.
\end{align}
The numerical similarity to $W_p(P_{33})$
suggests a close relationship between a dynamical coupled-channel
approach and the CM $K$-matrix parametrization.

It can be effectively argued that the essentially elastic nature of
the $P_{33}$ partial wave and the fact that nonresonant effects are
small in the region of the $P_{33}(1232)$ resonance are specific,
perhaps, only to this particular partial wave. We would not dispute
this claim. The purpose of the present comparison, however, is to
indicate that the dynamics of rescattering, including final-state
interactions and coupled-channel effects, are included in an essential
way into the CM parametrization. In the specific case of the $P_{33}$
partial wave, which provides a simple case study, there is a
quantitative verification of this fact given by
Eq.\eqref{eqn:barepole}.  We elaborate on this point further by
analyzing and comparing the rescattering effects in the model and
parametrization approaches.

There are two distinct sources of rescattering effects which dress the
explicit pole in the CM parametrization. The first, comes about in
Eq.\eqref{eqn:HKCMK}, where the real part of the Chew-Mandelstam
function, $\re C$ reflects the off-shell propagation of the
intermediate two-particle states. In fact, the subtraction constant
in $C$\cite{Paris:2010tz} reflects the model dependence of the
off-shell components. This off-shell rescattering effect dresses the
explicit pole in the CM $K$ matrix, shown in the first term of
Eq.\eqref{eqn:KCMme}, and yields the pole in the Heitler $K$ matrix
given in Eqs.\eqref{eqn:KpoleSAID} and \eqref{eqn:Kpoleval}. The
second source of rescattering effects in the CM parametrization form
induce a further shift of the Heitler $K$ matrix pole, $W^H_p(P_{33})$
to determine the position of the pole of the corresponding partial
wave $T$ matrix. The proximity of this pole to the physical region
dictates that the pole is, to the best of our knowledge, universally
found in model and parametrization treatments alike to be located at
$E=(1210,-51) \mev$ within 1 or 2 \mev.

A similar structure obtains for the Heitler $K$ matrix in the
dynamical model
\begin{align}
\label{eqn:KV}
K(E) &= V + V g(E) K(E),
\end{align}
where $g(E) = \re G_0(E)$ and we have suppressed a (continuous) sum
over intermediate, off-shell states in the second term. Since the bare
pole features in the resonant contribution to the interaction kernel
$V$ the off-shell rescattering effects in the second term will result
in a shift of the bare pole location, Eq.\eqref{eqn:barepole} to the 
Heitler $K$ matrix pole position. 

\begin{figure}
\includegraphics[width=8.5cm]{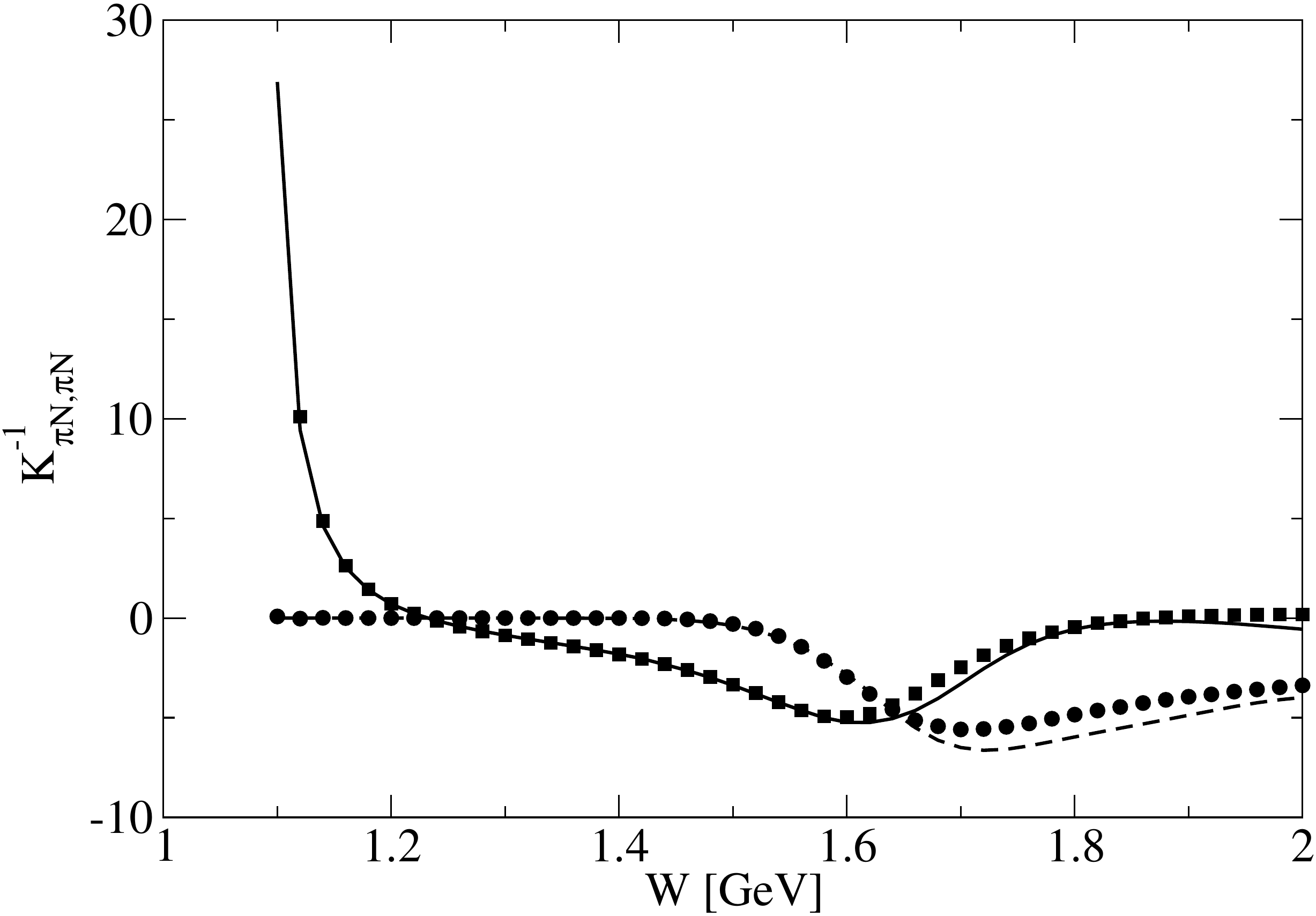}
\caption{\label{fig:invK}The inverse Heitler $K$ matrix element
$K^{-1}_{\pn,\pn}$ as function of the center-of-mass energy, $W$
calculated using Eq.\eqref{eqn:invKfig}. The dynamical model elements
of Ref.\cite{Paris:2008ig} are shown as continuous curves: solid
(dashed) curves show the real (imaginary) part.  The {\sc SP06}
solution of Ref.\cite{Arndt:2006bf}, which are very similar to the
dynamical model curves, are indicated by squares (circles) for the
real (imaginary) parts. The imaginary part, which should strictly be
zero by unitarity, deviates from zero above $W \approx 1.45$ GeV since
we have inverted the matrix element by ignoring inelastic effects. The
pole position of the Heitler $K$ matrix, Eq.\eqref{eqn:Kpoleval} may
be read off the figure at the node of $K^{-1}_{\pn,\pn}$.}
\end{figure}

Verification of the Heitler $K$ matrix pole positions in the dynamical
model and CM parametrization may be carried out graphically, as shown
in Fig.\eqref{fig:invK}. We may obtain the inverse of the $\pn\to\pn$
$K$-matrix element directly from the on-shell $\pn\to\pn$ $T$-matrix
element calculated in the dynamical model approximately as
\begin{align}
\label{eqn:invKfig}
K_{\pn,\pn}^{-1} &\approx i + T_{\pn,\pn}^{-1}.
\end{align}
for values of $W$ above the $\pn$ threshold.
Here we have absorbed the factor $\rho$ from Eq.\eqref{eqn:Heitler} by
a redefinition of $K$ and $T$. The zero of the inverse Heitler
$K$-matrix element, $K_{\pn,\pn}$ is the location of the matrix
element's pole. The derivation of Eq.\eqref{eqn:invKfig} has neglected
the other, inelastic channels.  The neglect of other channels, present
in both the dynamical model of Ref.\cite{Paris:2008ig}, which includes
$\pD,\rn,\sn,$ and $\en$ and the SAID {\sc SP06} parametrization,
which includes $\pD,\rn,$ and $\en$ is warranted, as shown in
Fig.\eqref{fig:invK}. In this figure, violations of unitarity incurred
by ignoring the inelastic processes appear as the imaginary part of
$K^{-1}_{\pn,\pn}$ deviates from zero, above about 1.45 GeV according
to the figure, where Eq.\eqref{eqn:invKfig} is no longer valid. The 
location of the pole in the Heitler $K$ matrix can
be read directly off the figure. The numerical value of $1.23$ GeV is
the same as that of the SAID parametrization given in
Eq.\eqref{eqn:Kpoleval}.

\section{Conclusion}
\label{sec:conc}
The dynamical coupled-channels approach has been compared with the CM
parametrization for the $P_{33}$ partial wave. The SAID 
parametrization, specifically the {\sc SP06} solution of
Ref.\cite{Arndt:2006bf}, includes an explicit pole in the CM $K$
matrix, $\Kbar$, as given in Eq.\eqref{eqn:KCMme}. The relationship
between the CM $K$ matrix and the Heitler $K$ matrix, $K$, depicted in
Eq.\eqref{eqn:HKCMK}, demonstrates that the `bare' CM pole is dressed
by off-shell effects, represented by the real part of the CM diagonal
matrix, $\re C(W)$. The Heitler $K$ matrix pole is further dressed by
the coupling to intermediate states in the continuum, through the
effects encoded in the relationship between the $K$ matrix and the $T$
matrix of Eq.\eqref{eqn:Heitler}.

Numerically, the description of the pole structure of $\Kbar$, $K$,
and $T$ bear a strong relationship to that of the dynamical model of
Refs.\cite{JuliaDiaz:2007kz} and \cite{Paris:2008ig}. In these models,
the bare pole of the interaction kernel, $V$ is related to the poles
of the $K$ matrix by Eq.\eqref{eqn:KV}. The second term of this
relation gives the rescattering effects of the off-shell intermediate
states, in close parallel to the form of Eqs.\eqref{eqn:HKCMK} or
\eqref{eqn:KinvReln}. Having established the numerical identity of the
Heitler $K$ matrix poles, the fact that the dynamical model is fit to
the $\pn$ elastic partial waves amplitudes\cite{JuliaDiaz:2007kz} one is
guaranteed that the $T$ matrix poles are virtually identical in both
the parametrization and model approaches.

The numerical identity of the pole structure of these two ostensibly
different approaches suggests a connection between them at the
dynamical level. The loop dynamics of the rescattering effects,
explicit in the microscopic, model dependent formulation of the
dynamical coupled channels approach is a fundamental aspect of
hadronic reactions and is believed to play a key role in
the understanding of single and multiple meson reactions. It is
encouraging that the CM approach, which gives model
independent\cite{Workman:2010ip} parametrizations of the $\pn$
elastic partial wave amplitudes, encodes the dynamics of the
model approach.

Finally, we emphasize in closing, that the specific value of the bare
pole in the dynamical model, $M^{(0)}(P_{33}) = 1391$ MeV is a model
dependent quantity. The value of this bare parameter, which appears in
the Lagrangian, depends on several factors including the assumed
unstable particle content (the assumed channel space), regularization
(\textit{ie.}, form factors) and approximations. In fact, the well
developed coupled-channel dynamical J\"{u}lich model gives a different
value for the bare mass. Reference \cite{Gasparyan:2003fp} gives, in
their Table (5.3), the value 1459 MeV. Indeed, model dependence can
even be seen within a given approach, if different model spaces and
approximations are considered. As examples of this, we can look at the
dynamical coupled-channel model of Ref.\cite{Sato:1996gk}, a precursor
of the models in Refs.\cite{JuliaDiaz:2007kz} and \cite{Paris:2008ig}.
Table I of Ref.\cite{Sato:1996gk} gives $M^{(0)}(P_{33})$ (their
$m_\Delta$) as 1299 (1319) MeV for model L(H). Alternatively, we
observe various values of the bare mass in different versions of the
J\"{u}lich model. Table I from Ref.\cite{Schutz:1994cp} we see the
value 1375 MeV, which can be compared to the value 1459 MeV quoted
from Ref.\cite{Gasparyan:2003fp} above. The origin of these shifts in
the bare pole mass are identifiable as a consequence of the particular
details of a given model formulation.  The shifts do not indicate, at
least to us, a measure of the quality of any given model.
Nevertheless, the model dependence of the bare parameters are seen
clearly and our earlier caveat appears warranted.  Here, however, we
have made the case that despite the model dependence of the bare pole
position, certain features of dynamical model treatements and our CM
parametrization approach are similar insofar as their dynamical
treatment is quantitatively comparable.

\begin{acknowledgments}
This work was supported in part by the U.S.\ Department of Energy
Grant DE-FG02-99ER41110.
\end{acknowledgments}

\bibliography{master}

\end{document}